# Reflectometry with registration of secondary radiation at total neutron reflection

V.D. Zhaketov[1], A.V. Petrenko[1], Yu.N. Kopatch[1], N.A. Gundorin[1], C. Hramko[1],
Yu.M. Gledenov[1], Yu.V. Nikitenko[1,*], V.L. Aksenov[1]

[1]*Joint Institute for Nuclear Research, Dubna, Russia*
*nikiten@nf.jinr.ru

Neutron reflectometry is a method for measuring of the spatial dependence (profile) of the potential interaction between neutron and medium. At interface of media the interaction potential is the sum of the element's potentials. For definition of potentials of separate elements (isotopes) a secondary radiation is recorded. Recording channels of secondary radiation are created on spectrometer REMUR at pulsed reactor IBR-2 in Dubna (Russia). The results for testing of the channels are reported and perspectives of reflectometry with registration of secondary radiation are discussed.

The properties of interface regions significantly change in a layered structure. So, for superconductor-ferromagnetic interface there is modification of magnetic spatial profile. At that a question is how relates the magnetic profile with element density profiles. Polarized neutron reflectometry [1] allows to measure the spatial profile of magnetic induction. But, the neutron-matter interaction potential (proportional to magnetic induction) is the sum of potentials of the interacting environment. So, the conventional reflectometry can not say what the elements are responsible for changing of a potential. For that a secondary radiation must be recorded [2]. Type of radiation and energy of one are signs which identify the isotopes of elements. Secondary radiation is charged particles[3], gamma-quanta and nuclear fission fragments. A more broad interpretation to secondary radiation follows to relate non-coherent scattering on nuclear, inelastic and diffuse neutron scattering. The special secondary radiation is the spin-flip neutrons.

For definition of position of isotope layer must be used special regimes of neutron field [4]. The first is standing waves, which forms as result of interference of initial and reflected waves. The second regime is enhanced standing waves which appears at interference of waves of different multiplicity neutron reflection from interfaces.

In the work the channels which are created on spectrometer REMUR at reactor IBR-2 in Dubna (Russia) are described. The results of testing of channels are reported and perspectives of reflectometry with registration of secondary radiation are discussed.

**Neutron absorption coefficient, model calculations.** Propagation of neutrons in medium at $k < \pi/d$, where $d$ is distance between of atoms, is described by optical potential (energy of potential interaction of neutron with medium) $U=V-iW$. For intensity of secondary radiation $J(k_z)$ has place

$$J(k_z) = \mu_d \iint \cos(\gamma_d)\cos(\gamma_s) \int v_z(z) n(z) N(z) \frac{d\sigma(z,\theta)}{d\Omega} dz d\Omega dS_d dS_s / S_d$$
$$\approx A \cdot B \cdot M \qquad (1)$$

where $A = \frac{\hbar}{m}\mu_d n_{\psi_0} k_{0z}\frac{\Delta\Omega}{4\pi}$; $B = \iint \cos(\gamma_d)\cos(\gamma_s)\, dS_d dS_s/S_d$; $M = \frac{j_{abs}}{j_0} = \int (n_\psi(z)k_z/n_{\psi_0}k_{0z})N(z)\sigma dz = n_\psi(z) = n_0|\psi|^2 = n_{\psi_0}(z) \cdot p^2$; $n_{\psi_0}(z) = n_0|\psi_0|^2$; $p(z) = \frac{\psi}{\psi_0}$, $k_w^2 = N\sigma k_z = \frac{2m}{\hbar^2}W$, $k_v^2 = \frac{2m}{\hbar^2}V$, $n_0$ and $\psi_0$ is initial neutron density and wave function, $n(z)$ and $\psi(z)$ is density and wave function in absorber layer, $U$ is interaction potential, $N$ is density of nuclear, $\sigma$ is cross-section of neutron with nuclear, $m$ is neutron mass, $\hbar$ is Plank constant, $S_d$, $S_s$ is area of detector and sample, $z$ is coordinate in depth structure, $\theta$ is grazing angle, $\Omega$ is solid angle, $v_z$ and $k_{0z}$ is perpendicular component of neutron velocity and wave vector, $\mu_d$ is efficiency of secondary radiation detector, $\gamma_s$, and $\gamma_d$ is direction angle of sample and detector.

From (1) it follows that $J(k_z)$ is proportional to absorption coefficient $M$. Absorption coefficient in turn is expressed through the $n_\psi(z)$ and $W(z)$. $W(z)$ is defined by nuclear density of element $N(z)$. A real part of potential on some orders exceeds the imaginary one. So, the $n_\psi(z)$ mainly is defined by sum of real parts of elements $\Sigma V_j \sim \Sigma(Nb)_i$. The sum $\Sigma V_j$ defines the reflection and transmission neutron coefficients also. The absorption coefficient in common case is equal of sum of partial absorption coefficients $M_{ij} \sim N_i \sigma_{ij}$, which correspond a nuclear type "$i$" and a type of secondary radiation "$j$".

So, a measurement of $R$, $T$ and $M_{ij}(J_{ij})$ allows to define the $\Sigma N_j(z)$ and $N_j(z)$. From it follows that in case of two isotopes a measurement of secondary radiation from one isotope only allows to define $N_j(z)$ for both isotopes. In case of three isotopes it necessary to measure secondary radiation from two isotopes and so on. Will demonstrate a sensitivity of measurements of the partial absorption coefficients. On fig. 1 it is shown model of structure, which has used at calculations. In this structure the bilayer is situated on distance $L$ from neutron reflector. Two types of interaction potentials are used for layers. The first the potentials have rectangle form. In second the potentials have triangle form and overlap on bilayer thickness. As result a thickness of the interface changes from minimal zero value to maximal one – thickness of bilayer. At that a quantity of matter does not change.

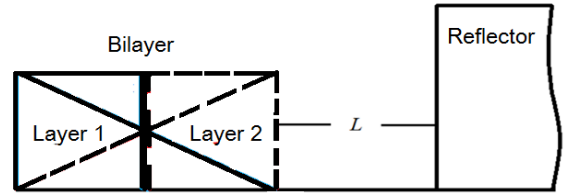

**Fig. 1.** Spatial dependence of interaction potential bilayer used at model calculations: potential of layers are rectangle or triangle form: layer 1 – solid line, layer 2 – dash line.

Fig. 2a presents dependence $M(k_z)$ for first (1,3) and second (2,4) layers with thickness 25 nm for rectangle (1,2) and triangle (3,4) an interaction potential at copper reflector. It is seen the dependences for two layers are significantly different. It reflects their the different position from reflector. Secondly a changing of interface form lead both to change of magnitude of maximums and their wave vector position.

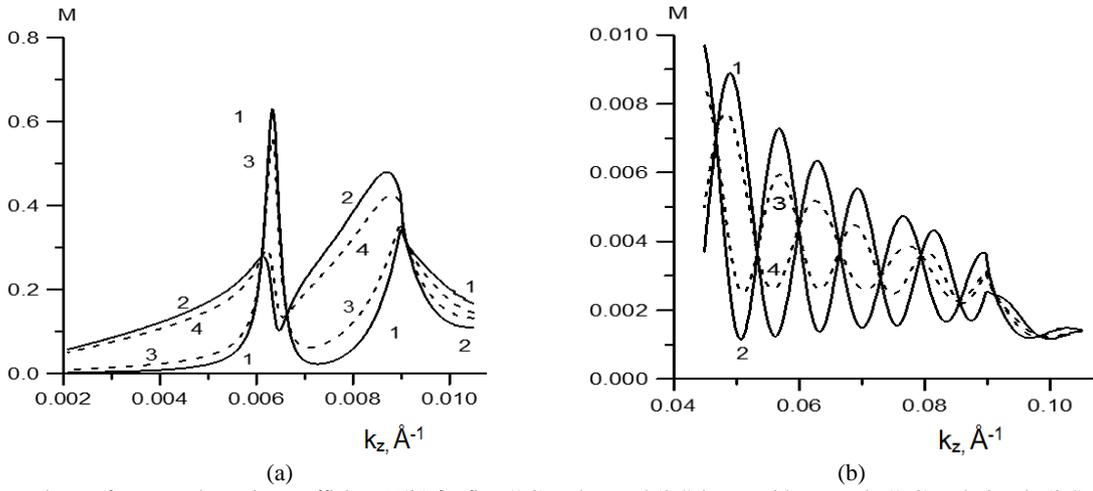

(a)                  (b)

**Fig. 2.** (a) Dependence of neutron absorption coefficient $M(k_z)$ for first (1,3) and second (2,4) layers with rectangle (1, 2) and triangle (3,4) potentials. For layers $k_v$ =0.007 Å$^{-1}$ and $k_w$ =0.0022 Å$^{-1}$, for reflector $k_v$ =0.009 Å$^{-1}$, bilayer thickness is 50 nm. (b) Dependence of neutron absorption coefficient $M(k_z)$ for first (1,3) and second (2,4) layers with rectangle (1, 2) and triangle (3,4) potentials. For layers $k_v$=0.007 Å$^{-1}$ and $k_w$=0.0022 Å$^{-1}$, for reflector $k_v$=0.09 Å$^{-1}$, bilayer thickness is 5 nm.

Fig. 2b presents similar to Fig. 2a dependences but for 5nm bilayer thickness and super mirror reflector for which critical real wave vector is 10 times more than for copper reflector. Obviously in this case a spatial resolution is in 10 times higher also.

We will consider now a case of polarized neutrons and magnetic structure. Next expression has place for absorption coefficient at initial neutron spin state "*i*" ("*f*" shows the second spin state)

$$M_i = \int \left( p_{ii}^2(z) + p_{if}^2(z) \right) k_w^2(z) dz / k_{0z} \quad (2)$$

where $p_{ii}(z) = \psi_i/\psi_{0i}, p_{if}(z) = \psi_f/\psi_{0i}$.

Fig. 3a shows dependences of spin flip neutron reflection coefficient (upper curves) and absorption coefficient (lower curves) at different values of $L$. It is seen that maxima of upper and lower curves coincide. It shows that $R_{sf}(k_z)$ can be used for defining the position of magnetic non-collinear layer also.

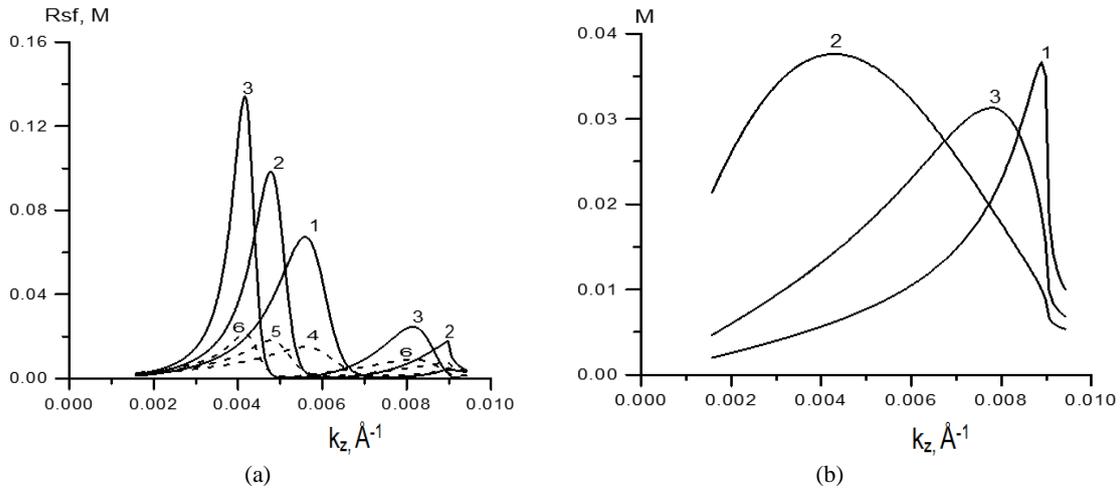

(a)                  (b)

**Fig. 3.** (a) Dependence of spin flip reflection coefficient (1-3) and absorption coefficient (4-6) for layer with thickness $h$=20 nm, $k_v$ =0.0053 Å$^{-1}$, $k_w$ =0.0002 Å$^{-1}$ and perpendicular to neutron polarization magnetization $J$=1 kOe, which is situated on distance $L$=20(1,4), 30(2,5) and 40 nm(3,6) from reflector with $k_v$ =0.009 Å$^{-1}$. (b) Dependence of absorption coefficient with initial "+" (1) and "-" (2) neutron polarization for layer with $h$=20 nm, $J$=10 kOe and $k_v$=0.0053 Å$^{-1}$, $k_w$=0.00053 Å$^{-1}$ and dependence of absorption coefficient for layer without of magnetization (3). Layer is situated at $L$=25 Å from reflector with $k_v$=0.009 Å$^{-1}$.

In the case of a collinear magnetic structure the dependence $M(k_z)$ (Fig. 3b) defines by initial neutron spin state. Obviously, a magnitude of magnetic induction defines the difference of the curves for plus (polarization $P$=+1) and minus ($P$=-1) neutron spin state.

**The channels for recording of the secondary radiation.**
Channels for recording of the secondary radiation are created at spectrometer polarized neutrons REMUR located at IBR-2 pulsed reactor in Dubna (Russia). IBR-2 reactor operates with a frequency of 5 Gz. Spectrometer REMUR has next basic parameters: distance from sample place to the moderator is 29 meters, distance from the sample place to neutron detector is 5 meters, the wavelength resolution at detector place is δλ=0.02 Å.

**The channel of polarized neutrons.** Neutron polarizer is super mirror $m$2 with a size of reflecting surface 10 cm×80 cm. Maximal solid angle of the polarizer from sample point is $\Omega_{max}$=2.0·10$^{-5}$ str. Neutron intensity of polarized neutrons at sample is 2·10$^4$ n/cm$^2$/s. Neutron polarization analyzer is a stack of 125 focused supermirrors $m$2. Entrance window of analyzer has size 18 cm×20 cm. On Fig. 4a it is presented the dependence of reflection intensity for different neutron spin transition for structure with a standing wave regime. It is seen that maxima of spin flip intensities corresponds to the minima of non-spin flip intensities. The dependence of the non-spin flip intensity is similar to the dependence of the neutron intensity for a non-magnetic structure.

The dependence of the spin flip intensity is similar to the dependence of the intensity of the secondary radiation.

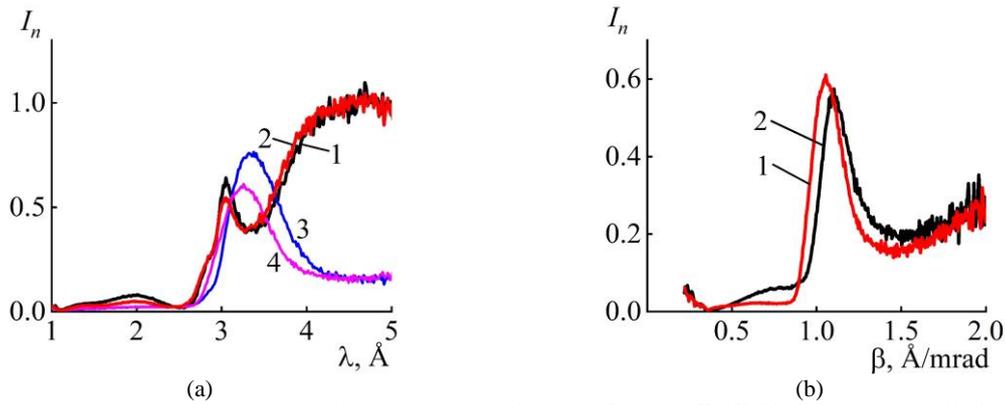

**Fig. 4.** (a) Reflection neutron intensities $I_n$ (1)-$I_n$ (4), which correspond states of two spin-flippers "off, off"(black curve), "on, on"(red curve), "off, on"(blue curve), "on, off"(сиреневая curve), for structure V(10nm)/CoFe(5nm)/$^6$LiF(5nm)/V(5nm)//glass at grazing angle θ=3 mrad, magnetic field $H$=295 Oe and magnetic field inclined angle φ=70°. (b) Spin flipped neutron intensities for structures V(20 nm)/CoFe(4 nm)/$^6$LiF(5 nm)/V(5 nm)//glass(1) and V(20 nm)/CoFe(4 nm)/$^6$LiF(5 nm)/V(15 nm)//glass(2).

Fig. 4b presents spin flip intensities for structures V(20 nm)/CoFe(4 nm)/$^6$LiF(5 nm)/V(5 nm)//glass(1) and V(20 nm)/CoFe(4 nm)/$^6$LiF(5 nm)/V(15 nm)//glass(2). The difference between dependencies 1 and 2 is because of different distance of the magnetic layer from the glass layer — the neutron reflector. It is seen, that with increasing of the distance from magnetic layer to the glass reflector the parameter β=λ/θ increases on 10%. From that follows that at λ=2 Å the minimal δ$L$ defined by uncertainty of neutron wavelength is 1 nm.

**The charged particle channel.** Ionization chamber was used for registration of the charged particles. Inside of the ionization chamber were placed investigated layered structures. Layered structures included neutron absorption layers $^6$Li$_{0.9}$$^7$Li$_{0.1}$F(5 nm).

Cross-section of thermal neutron capture by isotope $^6$Li is 945 barn. Result of reaction is an alfa-particle ($^4$He) and triton ($^3$H). On Fig. 5a and 5b is shown normalized on neutron intensity the recoded triton intensities for structures V(20 nm)/CoFe(4 nm)/$^6$LiF(5 nm)/V(5 nm)/glass(1), V(20 nm)/CoFe(4 nm)/$^6$LiF(5 nm)/V(15 nm)/glass(2) and Cu(10 nm)/V(65 nm)/CoFe(4 nm)/$^6$LiF(5 nm)/V(5 nm)/glass(1), Cu(10 nm)/V(55 nm)/CoFe(4 nm)/$^6$LiF(5 nm)/V(15 nm)/glass(2), correspondently

From Fig. 5a it follows that at δ$L$=10 nm the changing of β=λ/θ is 10%. As a result, minimal value δ$L$ is 1nm. δ$L$=1 nm is the same that in case of polarized neutrons.

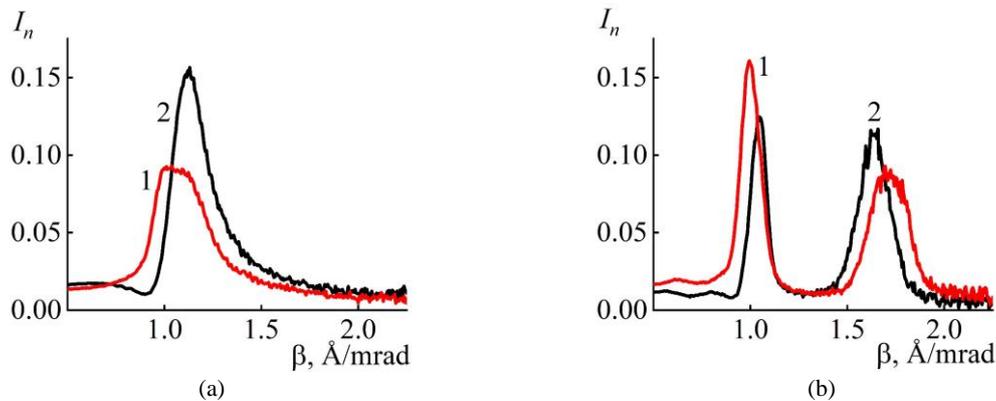

**Fig. 5.** (a) Experimental normalized intensities of charged particles for structures V(20 nm)/CoFe(4 nm)/$^6$LiF(5 nm)/V(5 nm)/glass(1) and V(20 nm)/CoFe(4 nm)/$^6$LiF(5 nm)/V(15 nm)/glass(2). (b) Experimental normalized intensities of charged particles for structures Cu(10nm)/V(65 нм)/CoFe(4 nm)/$^6$LiF(5 nm)/V(5nm)/glass(1) and Cu(10nm)/V(55 nm)/CoFe(4 nm)/$^6$LiF(5 nm)/V(15 nm)/glass(2).

From Fig. 5b follows that increasing of $L$ on 10 nm leads to decreasing the first resonance peak (β≈1 Å/mrad) on 22% and increasing of second peak (β≈1.7 Å/mrad) on 30%. It leads to the conclusion that for structure with enhanced standing wave regime a minimal δ$L$ is less (not more 0.5 nm) than for structure with standing wave regime.

**The gamma-quanta channel.** For the registration of gamma-quanta it was used a semiconductor germanium detector with efficiency 45% and energy resolution 2 keV for line 1.33 MeV. Detector have been situated on distance 40 mm from a sample. For testing of channel were used the structure with natural gadolinium 5 nm layers. Strongest transition in $^{157}$Gd at gamma-quanta energy 181.94 keV was used. Fraction of the transition in total cross-section is 18.33%. Prevalence of $^{157}$Gd in natural gadolinium is 15.68%. As result a cross-section for given transition in natural gadolinium at λ=1.8 Å was σ=7.3 kbarn. On Fig. 6a for example is shown gamma-quantum energy spectrum for structure V(20 nm)/Gd(5 nm)/V(5 nm)/Cu(100 nm)/glass.

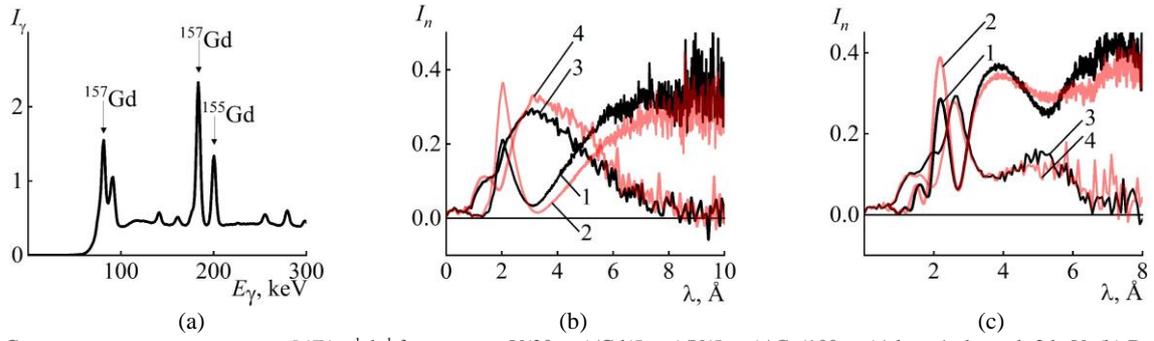

**Fig. 6.** (a) Gamma-quanta energy spectrum $I_\gamma(E_\gamma)$ s$^{-1}$ch$^{-1}$ for structure V(20 nm)/Gd(5 nm)/V(5 nm)/ Cu(100 nm)/glass, 1 channel=2 keV. (b) Dependences of normalized neutron reflection (1,2) and gamma-quanta (3,4) intensities for structures V(20 nm)/Gd(5 nm)/V(5 nm)/Cu(100 nm)/glass and V(10 nm)/Gd(5 nm)/ V(15 nm)/Cu(100 nm)/glass. (c) Dependences of normalized neutron reflection (1,2) and gamma-quanta (3,4) intensities for structures Cu(10 nm)/V(65 nm)/Gd(5 nm)/V(5 nm)/Cu(100 nm)/glass and Cu(10 nm)/ V(55 nm)/Gd(5nm)/V(15nm)/Cu(100nm)/glass.

Fig. 6b presents dependences of normalized neutron (1,2) and gamma-quanta (3,4) intensities for structures V(20 nm)/Gd(5 nm)/V(5 nm)/Cu(100 nm)/glass and V(10 nm)/Gd(5 nm)/V(15 nm)/Cu(100 nm)/glass. For $\lambda_1$=3.1 and $\lambda_2$=3.3 Å for corresponding structures were observed the minimums for dependences 1 and 2 and maximums for dependences 3 and 4. Difference $\Delta\lambda=\lambda_2-\lambda_1$=0.2 Å corresponds the 10 nm change of distance gadolinium layer from cupper reflector. As in case of polarized neutron channel and charged particles the minimal change of distance between absorption layer and reflector equals 1 nm.

On fig. 6c are presented the same dependences as on Fig. 6b but for structures Cu(10 nm)/V(65 nm)/Gd(5 nm)/V(5 nm)/Cu(100 nm)/glass and Cu(10 nm)/V(55 nm)/Gd(5 nm)/V(15 nm)/Cu(100 nm) /glass. In range of total reflection from cupper ( $\lambda > 2$Å ) at $\lambda$=2.7 and 5.15 Å two minima on neutrons curves and corresponding two maxima on gamma-quanta dependencies are observed. From the data follows that at $\lambda\approx1.5$ Å, where the neutron intensity maximal, the statistical accuracy of definition of absorber layer position is not more 0.5 nm.

In conclusion we will estimate the isotope minimal cross-section nuclear in layer which possible to investigate at REMUR. Will use a relation for measurement time at which the statistical error of secondary radiation count is equal to count

$$t = \frac{2J_{bgr}+J_\sigma}{J_\sigma^2} \quad (3)$$

where $\delta N = \sqrt{t \cdot J_{bgr}}$, $N_\sigma = t \cdot J_\sigma$, $J_{bgr}$ is background intensity.

We consider the next parameter values: measurement time $t$=1 day, wave vector resolution $\delta k/k$=0.1, neutron wavelength $\lambda$=1.5 Å, neutron beam cross-section on sample $S_{beam}$=0.1 cm$^2$, neutron flux at sample $J$=2×10$^4$ n/cm$^2$/s, thickness of absorber layer $h$=5 nm. With these parameters for channel of charged particle we have cross-section $\sigma_{min}$=0.025 barn. At that 22 isotopes are available for measurements. For gamma-quanta channel we have $\sigma_{min}$=0.3 barn, 91 isotopes and elements are available for investigations. In channel of polarized neutrons the minimal available for measurements a magnetization is 1 Gs.

**Conclusions**

In real time at the neutron spectrometer REMUR enough many isotopes are available for measurements with registration of secondary radiation. Further progress is connected with increasing of neutron intensity in 10 times, decreasing of the background of fast neutrons and gamma radiation from active reactor zone in 4-10 times, increasing of gamma detector solid angle in 4-5 times. Together using of these measures will allows to reach $\sigma_{min}\approx$1 mbarn at $h$=5 nm or $h\approx$1 Å at $\sigma_{min}$ =50 mbarn. With supermirror reflector can be achieved a spatial resolution 1 Å also.